\documentclass{iau} 
\usepackage{graphicx}
\usepackage{hyperref}        

\def\lum {\mbox{erg\,s$^{-1}$}}

\title[The neutron star zoo] 
{Exploring the neutron star zoo: \\ An observational review}

\author[A. Borghese]   
{Alice Borghese$^{1,2}$}

\affiliation{$^{1}$Institute of Space Sciences (ICE, CSIC), Campus UAB, Carrer de Can Magrans s/n, E-08193, Barcelona, Spain\\ [\affilskip]
$^{2}$Institut d'Estudis Espacials de Catalunya (IEEC), Carrer Gran Capit\`a 2--4, E-08034 Barcelona, Spain\\
email: {\tt borghese@ice.csic.es}}

\pubyear{2022}
\volume{363}  
\jname{Neutron Star Astrophysics at the Crossroads: \\ Magnetars and the Multimessenger Revolution}
\editors{E. Troja \& M. Baring, eds.}

\begin{document}

\maketitle

\begin{abstract}
Neutron stars have shown diverse characteristics, leading us to classify them into different classes. In this proceeding, I review the observational properties of isolated neutron stars: from magnetars, the strongest magnets we know of, to central compact objects, the so-called anti-magnetars, stopping by the rotation-powered pulsars and X-ray dim isolated neutron stars. Finally, I highlight a few sources that have exhibited features straddling those of different groups, blurring the apparent diversity of the neutron star zoo.
\end{abstract}

\firstsection 

\section{Introduction}

Remnants of massive stars, neutron stars (NSs) come in different flavours. Historically, we classify them according to the primary source of energy that feeds their emission. For instance, rotation-powered pulsars (RPPs) are driven by the loss of rotational energy due to the braking caused by their magnetic field, while accretion-powered NSs acquire their energy from matter transferred from a companion star to the stellar surface (their emission is due to the release of gravitational binding energy of the accreting material). Another possible energy reservoir is the internal heat, which can be either the results of surface reheating from external sources or the leftover from the star formation. Finally, the decay and instability of a super strong magnetic field (up to 10$^{15}$\,G at the surface) can power the persistent emission and bursting episodes of the so-called magnetars, the most magnetised NSs we know of in the Universe. These different mechanisms define different NS classes, such as the already mentioned RPPs and magnetars, besides the thermally-emitting X-ray dim isolated neutrons stars (XDINSs), the central compact objects (CCOs) and the millisecond pulsars (MSPs) in binary systems. In this proceeding, I will focus on {\it isolated} NSs, i.e. any NS that is not accreting matter. The remaining three energy sources are not mutually exclusive and, indeed, we witnessed phenomena that blur the boundaries between the different groups and point towards the possibility of a {\it grand unification} of the diverse observational manifestations of isolated NSs.\\  

\begin{figure}[ht]
\begin{center}
\includegraphics[width=0.8\columnwidth]{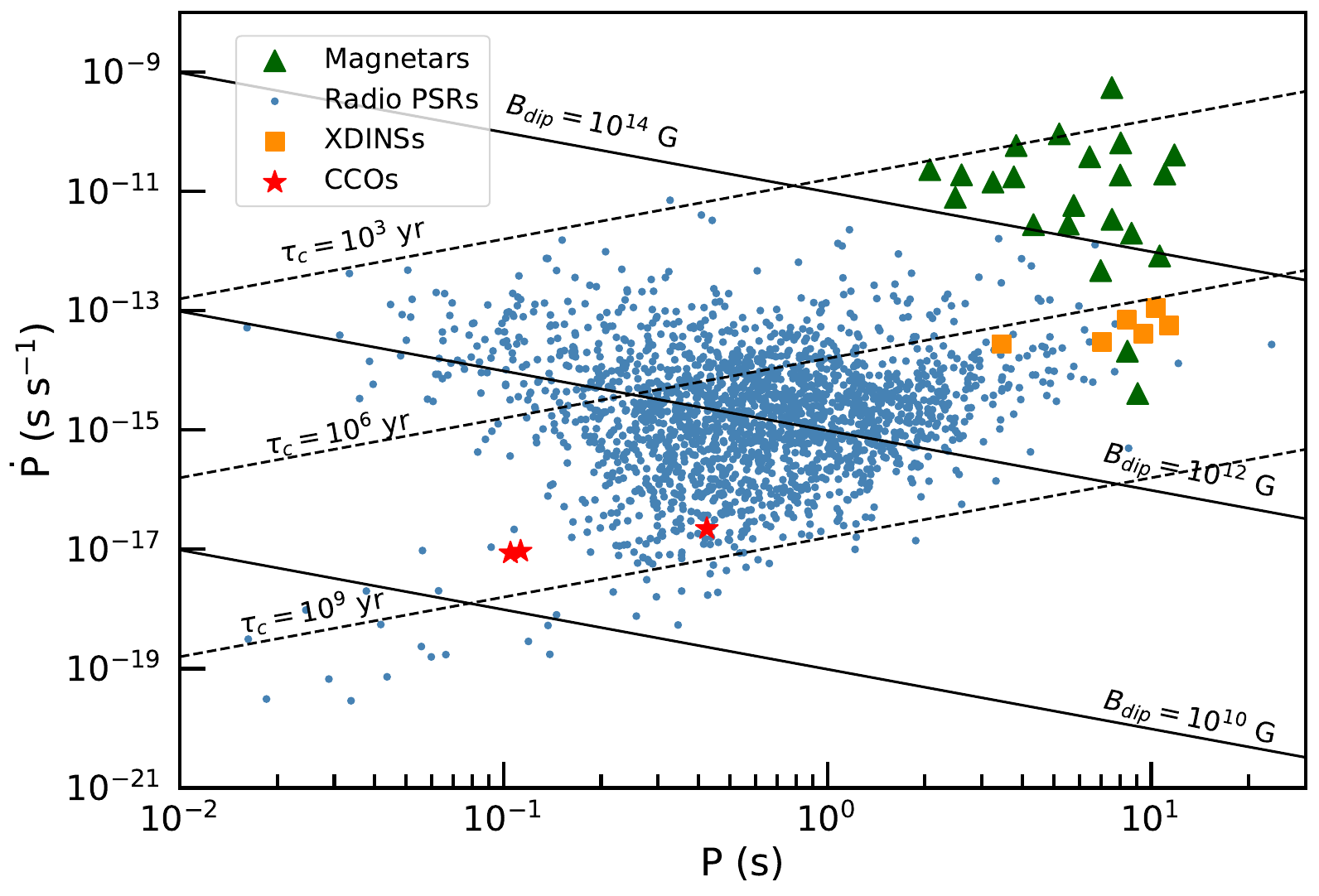} 
 \caption{$P$--$\dot{P}$ diagram with the isolated neutron stars included in the ATNF pulsar catalogue v.1.65. Green triangles, blue dots, orange squares and red stars represent magnetars, rotation-powered pulsars, X-ray dim isolated neutron stars and central compact objects, respectively. Also plotted are lines of constant dipolar magnetic field (solid lines) and characteristic age (dashed lines) as derived from equations \ref{eq:Bdip} and \ref{eq:tauc}. Image credit: A. Borghese.}
\label{fig:ppdot}
\end{center}
\end{figure}

The spin period $P$ and its first derivative $\dot{P}$ are two of the primary observables. The $P$--$\dot{P}$ diagram (Figure\,\ref{fig:ppdot}) provides the easiest way to visualise the isolated NS population. From a measurement of $P$ and $\dot{P}$ of an isolated NS, we can derive an estimate of the dipolar component of the magnetic field at the NS surface. According to the magnetic-dipole braking model (see e.g., \cite[Pacini (1967)]{1967Natur.216..567P}), a NS is considered as a rotating magnetic dipole in vacuum. By assuming that the NS spin down is dominated by the magnetic field torque, that the emission process is dipole radiation and that the neutron star is an orthogonal rotator, the strength of the dipolar magnetic field at the equator is equal to
\begin{equation}
    B_{\rm dip} \simeq 3.2 \times 10^{19} (P \dot{P})^{1/2}\; \rm{G}, 
    \label{eq:Bdip}
\end{equation}
where standard parameters for a NS have been considered (mass $M=1.4M_\odot$ and radius $R=10$\,km). A general spin down formula, $\dot{\nu} \propto \nu^{n}$, can be derived from the above equation, where $\nu=P^{-1}$ is the spin frequency and $n \propto \nu \ddot{\nu} \dot{\nu}^{-2}$ is the braking index, related to the slow down mechanism (e.g., $n = 3$ for pure magnetic dipole braking). By integrating the spin down formula, we obtain the NS age
\begin{equation}
    \tau = \frac{P}{(n-1) \dot{P}} \left[ 1 - \left( \frac{P_0}{P} \right)^{n-1} \right],
    \label{eq:tau}
\end{equation}
where $P_0$ is the spin period at birth. Under the assumption that the spin period at birth is much shorter than the current value and that the spin-down is due to magnetic dipole radiation ($n = 3$), $\tau$ no longer depends on $P_0$ and equation\,\ref{eq:tau} simplifies to the characteristic age
\begin{equation}
    \tau_{\rm c} = \frac{P}{2 \dot{P}}.
    \label{eq:tauc}
\end{equation}
Note that $\tau_{\rm c}$ does not necessarily provide a reliable estimate of the true age of the NS because of the many assumptions and simplifications, but for most pulsars is the only one available.

Although the magnetic-dipole braking model assumes that the star rotates in vacuum, which is not realistic, it yields estimates of important physical parameters just with a measurement of $P$ and $\dot{P}$. For this reason, the $P$--$\dot{P}$ diagram is a useful tool. Figure\,\ref{fig:ppdot} shows the $P$--$\dot{P}$ phase space with the isolated NSs included in the Australia Telescope National Facility (ATNF) Pulsar Catalogue, version 1.65\footnote{\url{https://www.atnf.csiro.au/research/pulsar/psrcat/index.html}, \cite[Manchester \etal\ (2005)]{2005AJ....129.1993M}.}. Lines of constant $B_{\rm dip}$ and $\tau_{\rm c}$ are also drawn. The RPPs  (blue dots) represent the bulk of the known NS population, with dipolar magnetic fields spanning $B_{\rm dip} \sim 10^{11}-10^{13}$\,G and characteristic ages between $\tau_{\rm c} \sim 10^{3}-10^{8}$\,yr. The CCOs (red stars) blend with the radio pulsars, but their dipolar fields are lower ($B_{\rm dip} \sim 10^{10}$\,G). In the top right corner, there are two groups of NSs, the magnetars (green triangles) and the XDINSs (orange squares), with $\tau_{\rm c} \sim 10^3-10^6$\,yr and $B_{\rm dip} \sim 10^{13}-10^{15}$\,G. In the following, I will discuss the observational features of these isolated NS classes in some detail.

\section{Magnetars}

At the time of writing (2022 January), the magnetar class counts 29 confirmed members\footnote{An update list is maintained at \url{http://www.physics.mcgill.ca/~pulsar/magnetar/main.html}.}, all residing in our Galaxy at low latitudes in the Galactic plane except for two sources located in the Magellanic Clouds. Magnetars show pulsations at relatively long periods ($P \sim 1-12$\,s), which are slowing down on time scales of a few thousand years. However, recently, a couple of detections for ultra-long period magnetars have been claimed (see contributions by Hurley-Walker and Stappers, this volume). The timing parameters yield characteristic ages $\tau_{\rm c} \sim 10^3 - 10^5$\,yr and external dipolar magnetic fields $B_{\rm dip} \sim 10^{14} - 10^{15}$\,G. Magnetars are characterised by persistent X-ray luminosity in the range $L_{\rm X} \sim 10^{31}-10^{36}$\,\lum, generally larger than the rotational energy loss rate. For this reason, the emission of these objects is thought to be ultimately powered by the decay and instabilities of their strong magnetic field (see e.g., \cite[Esposito \etal\ (2021)]{2021ASSL..461...97E} for a review).      

Magnetars are observed to emit over a broad energy range, from soft X-rays up to $\sim$150\,keV (e.g., \cite[G{\"o}tz \etal\ (2006)]{2006A&A...449L..31G}). The soft X-ray emission is described by a thermal component (a blackbody with temperature $kT \sim 0.3-1$\,keV) and, in some cases, a second component is required, either a hotter blackbody ($kT \sim 1-2$\,keV) or a power law with photon index $\Gamma \sim 2-4$. This decomposition is generally interpreted in a scenario according to which the thermal emission arises from the hot NS surface and can be distorted by magnetospheric effects, such as resonant cyclotron scattering (for a detailed discussion about these models see \cite[Turolla \etal\ (2015)]{2015RPPh...78k6901T}). The hard X-ray component is adequately modelled by a power-law spectrum flatter than that observed in the soft X-rays, with photon index $\Gamma \sim 0.5-2$. The mechanism at the origin of these hard tails is still under debate. The prime candidate invokes resonant Compton up-scattering of soft thermal photons by a population of highly relativistic electrons threaded in the magnetosphere (see e.g., \cite[Wadiasingh \etal\ (2018)]{2018ApJ...854...98W}).    

The hallmark of magnetars is the unpredictable and variable bursting activity observed in the X-/$\gamma$-ray regime on different timescales. Ranging from milliseconds to several tens of seconds, three kinds of flaring events have been observed:

1. {\it Giant flares}: So far, only three giant flares from three different Galactic magnetars have been recorded reaching peak luminosity of $\sim 10^{44}-10^{47}$\,\lum. All three events showed an initial brief ($\sim 0.1-0.2$\,s) spike of $\gamma$-rays, followed by a hard X-ray tail modulated at the NS spin period and observed to decay in a few minutes;

2. {\it Intermediate bursts}: They are characterised by a duration spanning between $\sim 1-40$\,s and have a peak luminosity ranging from $10^{41}$\,\lum\ and $10^{43}$\,\lum. They are marked by an abrupt onset and usually show thermal spectra;

3. {\it Short bursts}: these are the most common events, with typical duration of $\sim0.01-1$\,s and peak luminosity of $\sim 10^{39}-10^{41}$\,\lum, and represent the main way to discover new magnetars. They can occur sporadically or clustered in time. 

\begin{figure}[ht]
\begin{center}
\includegraphics[width=0.9\columnwidth]{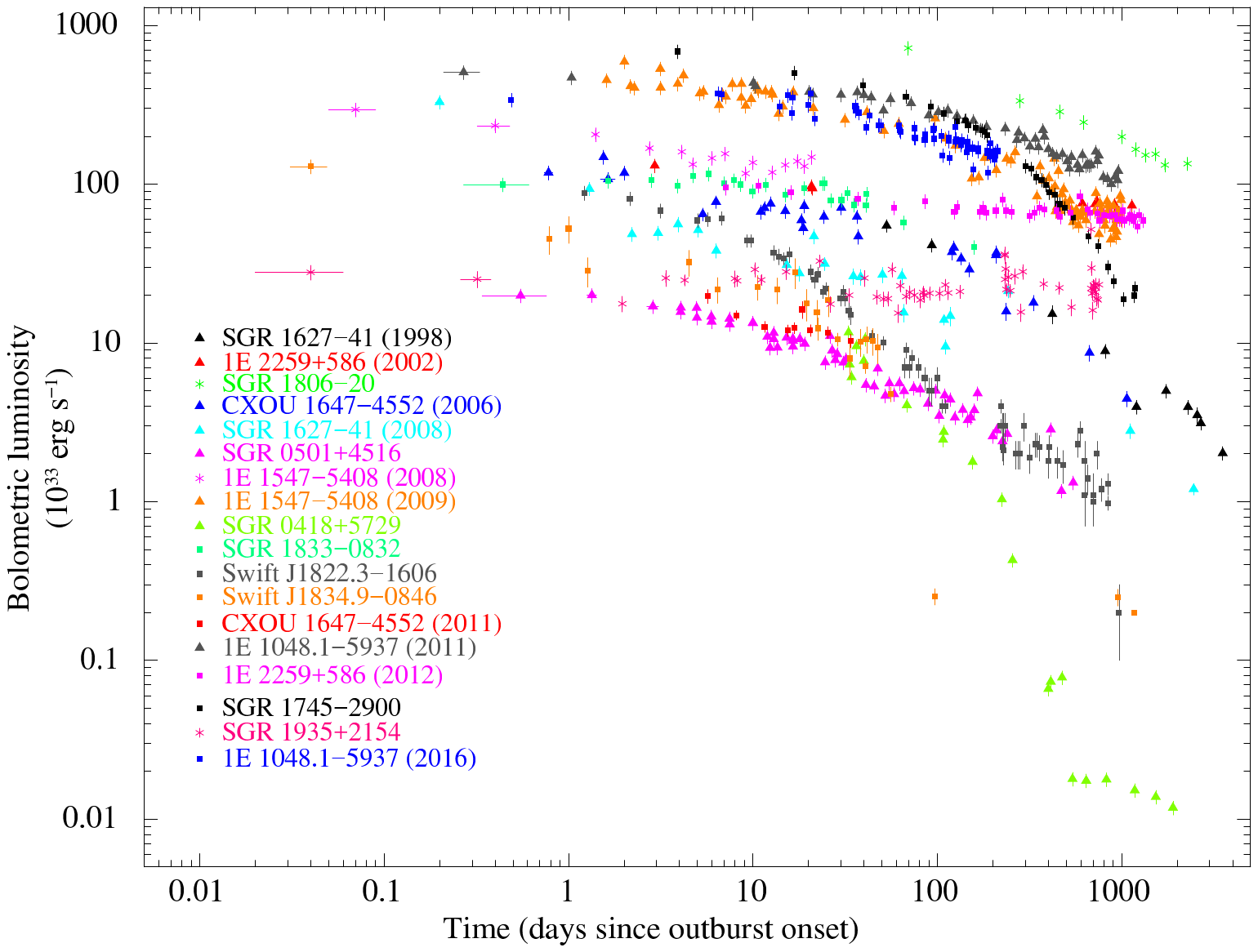} 
 \caption{Temporal evolution of the bolometric ($0.01-100$\,keV) luminosities for the major outbursts occurred up to the end of 2016 and with good observed coverage. Image from \cite[Coti Zelati \etal\ (2018)]{2018MNRAS.474..961C}.}
\label{fig:MOOC}
\end{center}
\end{figure}

The bursts often announce that a source has entered an active phase, commonly referred to as {\it outburst}. During an outburst, the persistent X-ray luminosity suddenly increases up to three orders of magnitude higher than the pre-outburst level. Then, it usually relaxes back to the quiescent level on timescales spanning from weeks to months/years (see Figure\,\ref{fig:MOOC} and the Magnetar Outburst Online Catalog at \url{http://magnetars.ice.csic.es/#/welcome}, \cite[Coti Zelati \etal\ (2018)]{2018MNRAS.474..961C}). During an outburst, the soft X-ray spectrum undergoes an overall initial hardening and then slowly softens on the timescales of the flux relaxation. In some cases, a transient non-thermal hard power-law tail is detectable (\cite[Enoto \etal\ (2017)]{2017ApJS..231....8E}). Outbursts are generally accompanied by timing anomalies, such as glitches or pulse profile changes, and by pulsed radio emission in six cases (see contribution by Possenti, this volume). It is believed that outbursts are caused by heat deposition in a restricted area of the magnetar surface, however the responsible heating mechanism is still poorly understood. Outburst activity is triggered by magnetic stresses in a localised region of the NS crust and these stresses deform part of the crust in a plastic way (see e.g., \cite[Li et al. (2016)]{2016ApJ...833..189L}, \cite[Gourgouliatos \& Lander (2021)]{2021MNRAS.506.3578G}). The plastic flows convert mechanically the magnetic energy into heat, which is conducted up to the surface layers and radiated in the form of thermal emission. Meantime, the plastic flows lead to thermoplastic waves that can move the crust. Crustal displacements can implant a strong twist of the magnetic field lines in the magnetosphere, most likely in a confined region. Additional heating of the surface layers is then produced by the currents flowing in the bundle as they hit the NS surface. Both mechanisms are probably at work during an outburst (see e.g., \cite[Beloborodov (2009)]{2009ApJ...703.1044B}, \cite[Beloborodov \& Li (2016)]{2016ApJ...833..261B}, \cite[De Grandis \etal\ (2020)]{2020ApJ...903...40D}).       

In the last decade, the belief that magnetars possess a super-strong magnetic field has been challenged by the discovery of three sources with a dipolar magnetic field within the range of those of ordinary radio pulsars (see \cite[Turolla \& Esposito (2013)]{2013IJMPD..2230024T} for a review and reference therein; \cite[Rea \etal\ (2014)]{2014ApJ...781L..17R}). SGR\,0418+5729, Swift\,J1822--1606 and 3XMM\,J185246.6+003317 are similar to the other magnetars (e.g., they emit bursts and underwent outbursts), apart from the strength of $B_{\rm dip}$ derived from the spin parameters, $B_{\rm dip} \sim (0.6-4) \times 10^{13}$\,G. Their characteristic age is about 2--3 orders of magnitude higher than the typical values for magnetars, suggesting that these sources are old magnetars where the magnetic field had already decayed. Another observational gauge of the magnetic field strength is provided by cyclotron features. Such features were reported for two low-$B$ magnetars (\cite[Tiengo \etal\ (2013)]{2013Natur.500..312T}, \cite[Rodr{\'\i}guez Castillo \etal\ (2016)]{2016MNRAS.456.4145R}). Given the strong dependence of the line on the NS rotational phase -- the line is detected only in a limited interval of the phase cycle, the most plausible explanation is proton cyclotron resonant scattering in a confined magnetic loop close to the NS surface. These discoveries are the first observational evidence for small-scale strong magnetic field structures, besides the dipolar component.  

\section{X-ray Dim Isolated Neutron Stars}

Up to now there are seven confirmed X-ray Dim Isolated Neutron Stars (XDINSs), for this reason they got the nickname of `The Magnificent Seven' (see \cite[Turolla (2009)]{2009ASSL..357..141T} for a review). All discovered by the {\it ROSAT} satellite, XDINSs are among the closest NSs we know of, with distances $\leq 500$\,pc. They are radio-silent, X-ray emitters with optical and/or ultraviolet counterparts. Timing studies revealed X-ray pulsations at spin periods in the range $3-12$\,s with period derivatives of the order of $10^{-14}-10^{-13}$\,s\,s$^{-1}$ for six of them, implying dipolar magnetic fields $B_{\rm dip} \sim 1-4 \times 10^{13}$\,G and characteristic ages $\tau_{\rm c} \sim 1-4$\,Myr (see Table\,\ref{tab:xdins_list}). XDINSs are characterised by soft X-ray ($0.1-1.5$\,keV) spectra modelled by an absorbed blackbody with low value for the absorption column density ($N_{H} \sim 10^{20}$\,cm$^{-2}$) and temperatures in the range $50-100$\,keV, without any evidence for the hard power-law component observed in other isolated NSs. 

Although a blackbody model provides an overall good description of the spectral energy distribution, broad absorption features had been found in most of the XDINSs. The properties of the spectral features are similar in the different sources: the central energies range from $\sim 300$\,eV to $\sim 800$\,eV, the equivalent widths are several tens of eV ($30-150$\,eV) and they appear to vary with the spin phase. Deviations from a pure blackbody can be explained by several physical mechanisms, such as cyclotron resonances in a magnetised atmosphere and atomic transitions in the outermost layers of the NS (\cite[van Kerkwijk \& Kaplan (2007)]{2007Ap&SS.308..191V}), or might be induced by an inhomogeneous surface temperature distribution (\cite[Vigan{\`o} \etal\ (2014)]{2014MNRAS.443...31V}).   

\begin{table}[ht]
\begin{center}
\caption{Overall properties of XDINSs. $E_0$ refers to the central energies of the broad absorption line. $B_{\rm dip}$ corresponds to the surface, dipolar strength of the magnetic field measured at the equator and $B_{\rm cyc}$ is the magnetic field strength evaluated assuming that the phase-average features are proton cyclotron resonances. $L$ indicates the bolometric luminosity. Adapted from \cite[Pires \etal\ (2014)]{2014A&A...563A..50P}.}
\label{tab:xdins_list}

\footnotesize{
\begin{tabular}{@{}lcccccccc}
\hline

Source & $kT$  & $P$ & $\log{\dot{P}}$ & $\tau_{\rm c}$ & $E_0$ & $B_{\rm dip}$ 	&  $B_{\rm cyc}$ & $\log{L}$   \\ 
	   & (eV)  & (s) & 	   (\ss)      & 	     (10$^6$~yr)       & (eV)  & (10$^{13}$~G) & (10$^{13}$~G) & (\lum) \\
\hline
RX\,J1856.5--3754 & 61     & 7.06 & -13.527  & 3.8 & --  & 1.47 & -- & 31.5-31.7   \\
RX\,J0720.4--3125$^a$ & 84--94 & 8.39 & -13.156  & 1.9 & 311 & 2.45 & 5.62 & 32.2-32.4 \\
RX\,J1605.3+3249$^b$  & 100    & $\dots$ & $\dots$  & $\dots$ & 400 & $\dots$ & 8.32 & 31.9-32.2 \\
RX\,J1308.6+2127  & 100    & 10.31 & -12.951 & 1.4 & 390 & 3.48 & 3.98 & 32.1-32.2 \\
RX\,J2143.0+0654  & 104    & 9.43 & -13.398  & 3.7 & 750 & 1.95 & 1.41 & 31.8-31.9 \\
RX\,J0806.4--4123 & 95     & 11.37 & -13.260 & 3.2 & 486 & 2.51 & 9.12 & 31.2-31.4 \\
RX\,J0420.0--5022$^c$ & 48     & 3.45 & -13.553  & 1.9 & $\dots$ & 1.00 & $\dots$ & 30.9-31.0 \\  
\hline

\end{tabular}}
\end{center}
$^a$ \cite[Hambaryan \etal\ (2017)]{2017A&A...601A.108H} claimed that the most-likely genuine period is 16.78\,s, twice that reported in the literature.\\
$^b$ A period of 3.39\,s was claimed but not confirmed by later observations.\\
$^c$ An absorption line at $\sim0.3$\,keV was reported, but not confirmed.
\end{table}

According to magneto-thermal evolutionary models (\cite[Vigan{\`o} \etal\ (2013)]{2013MNRAS.434..123V}), XDINSs are believed to be old magnetars; therefore, signature of a stronger magnetic field is expected to be found. This was the case for two XDINSs, RX\,J0720.4--3125 and RX\,J1308.6+2127, where narrow phase-dependent absorption features were discovered with characteristics similar to those of the lines reported in the low-$B$ magnetars (\cite[Borghese \etal\ (2015)]{2015ApJ...807L..20B}, \cite[Borghese \etal\ (2017)]{2017MNRAS.468.2975B}). In both sources, the feature is detected in only 20\% of the star rotational phase and the line energy is about $\sim0.75$\,keV. Due to the similarities with the detections in the low-$B$ magnetars, the spectral line might be explained by invoking the same physical mechanism: proton cyclotron resonant scattering in a localised loop close to the NS surface with magnetic field of about a factor of $\sim 5$ higher than the dipolar component. These findings strengthen the evolutionary link between magnetars and XDINSs, and, moreover, support the picture in which the magnetic field of highly magnetised NS is more complex than a pure dipole with deviations on a small scale, as expected from simulations.  

\section{Rotation-powered Pulsars and exceptions}

Rotation-powered Pulsars (RPPs) represent the bulk of the isolated NS zoo (see e.g., \cite[Kaspi \& Kramer (2016)]{2016arXiv160207738K} for a review). The known RPPs population is currently consisting of over 2800 sources with numbers constantly increasing thanks to ongoing radio surveys. They rotate rapidly with periods $P$ of a few hundred milliseconds and spin down steadily with a $\dot{P}$ typically of tens of microseconds per year. The key physical properties $B_{\rm dip}$ and $\tau_{\rm c}$ inferred from the timing parameters range between $10^{11}-10^{13}$\,G and $10^{3}-10^{8}$\,yr, respectively. RPPs are mainly observed in the radio band. However, they emit across the electromagnetic spectrum. Many sources have been detected in the X-ray band, where the emission can be thermal or non-thermal. The first arises from the surface and is characterised by broad pulses; the latter originates in the magnetosphere and appears highly beamed. Moreover, several RPPs have been observed in $\gamma$-rays with emission being non-thermal (see the second {\it Fermi} Large Area Telescope catalog; \cite[Abdo \etal\ (2013)]{2013ApJS..208...17A}). The $\gamma$-ray radiation comes from the magnetosphere, where charged particles are accelerated along the magnetic field lines. It still an open question where exactly the acceleration takes place, two main models have been put forward (see e.g., \cite[Harding (2021)]{2021arXiv210105751H} for a review). According to polar cap models, the particles are accelerated above the neutron star surface and the $\gamma$-ray emission is due to curvature radiation or inverse Compton. On the other hand, outer gap models assume that the acceleration occurs outer in the magnetosphere and $\gamma$-rays result from photon-photon pair production-induced cascades.

There are a few objects that stands out from the RPP typical profile:

$\bullet$ {\it Slow Radio Pulsars}: in the last years, a couple of slow ($P > 10$\,s) pulsars have been discovered. The radio pulsar PSR\,J0250+5854 was discovered on 2017 July 30 by the LOFAR Tied-Array All-Sky Survey (\cite[Tan \etal\ (2018)]{2018ApJ...866...54T}). It has a remarkably long spin period of $P \sim 23.5$\,s, the longest observed so far for a radio pulsar. Inspection of earlier observations yielded a detection allowing to infer a pulsar spin-down rate of $\dot{P} \sim 2.7 \times 10^{-14}$\,s\,s$^{-1}$. The observed spin parameters indicate $B_{\rm dip} \sim 2.6 \times 10^{13}$\,G and $\tau_{\rm c} \sim 13.7$\,Myr.Secondly, PSR\,J2251--3711 was found in the SUrvey for Pulsars and Extragalactic Radio Bursts conducted at Parkes (\cite[Morello \etal\ (2020)]{2020MNRAS.493.1165M}). It has a period of $P \sim 12.1$\,s and a spin-down rate of $\dot{P} \sim 1.3 \times 10^{-14}$\,s\,s$^{-1}$, implying $B_{\rm dip} \sim 1.3 \times 10^{13}$\,G and $\tau_{\rm c} \sim 14.7$\,Myr. For both pulsars, the timing properties are peculiar with respect to those measured for the majority of radio pulsars. The slow rotational period and relatively high magnetic field place these objects close to the XDINSs and old magnetars in the $P$--$\dot{P}$ diagram, bridging RPPs with these two other NS groups; 
 
$\bullet$ {\it High-$B$ Radio Pulsars}: there is a small group of RPPs with magnetar field strengths ($B_{\rm dip} \geq 10^{13}$\,G; see e.g., \cite[Ng \& Kaspi (2011)]{2011AIPC.1379...60N} for a review). Given that the strong magnetic field drives the explosive events in magnetars, magnetar-like activity might be expected from high-$B$ pulsars and, indeed, has been observed in two such sources. Associated with the supernova remnant Kes\,75, PSR\,J1846$-$0258 is one of the youngest known pulsars in the Galaxy and has a $B_{\rm dip} \sim 5 \times 10^{13}$\,G. It behaves as a canonical RPP for most of its observed lifetime, with X-ray luminosity lower than its spin down luminosity and powering a bright pulsar wind nebula. However, no radio emission has been detected so far. In May 2006, the source emitted a few X-ray magnetar-like short bursts, accompanied by a sudden increase of the flux that lasted about two months (\cite[Gavriil \etal\ (2008)]{2008Sci...319.1802G}). The spectrum underwent from a purely magnetospheric-type, typical of energetic RPPs, to one consistent with the persistent emission of magnetars. After 14 years of quiescence, PSR\,J1846$-$0258 entered again in outburst in 2020 August (\cite[Blumer \etal\ (2021)]{2021ApJ...911L...6B}). The second high-$B$ pulsar showing a magnetar outburst is PSR\,J1119$-$6127, at the centre of the supernova remnant G292.2$-$0.5 and with a $B_{\rm dip} \sim 4 \times 10^{13}$\,G (see also the contribution by Safi-Harb, this volume). In 2016 July, two bursts marked the onset of a magnetar-like outburst (\cite[Archibald \etal\ (2016)]{2016ApJ...829L..21A}, \cite[G{\"o}{\u{g}}{\"u}{\c{s}} \etal\ (2016)]{2016ApJ...829L..25G}). A flux enhancement by a factor of $> 160$ was registered, and the spectrum underwent a hardening with temperature increasing from $\sim 0.2$\,keV to $\sim 1.1$\,keV. At the outburst onset, the pulsar turned off as a radio-loud source and after two weeks it displayed a magnetar-like radio spectral flattening.

The study of these kind of RPPs that bridge different isolated NS group is fundamental to get an overall picture of the NS zoo.

\section{Central Compact Objects}

Central Compact Objects (CCOs) form a small group of young X-ray emitting isolated NSs, observed close to the centre of supernova remnants and characterised by the absence of emission at other wavelengths (see e.g., \cite[De Luca (2017)]{2017JPhCS.932a2006D} for a review). This class consists of a dozen of confirmed sources that show thermal spectra in the soft X-ray range (0.2--0.5\,keV) with no evidence for non-thermal emission. A single blackbody does not provide a satisfactory description in most cases, calling for a second blackbody component. The inferred temperatures are $\sim 0.2-0.5$\,keV for the colder component and about a factor of $\sim2$ higher for the hotter blackbody; the emitting regions are very small with radii that range from 0.1 to a few km. Pulsations in the $0.1-0.4$\,s interval have been detected only for three CCOs, confirming that they are, indeed, NSs, and long observational campaign allowed to measure the corresponding period derivatives of the order of $10^{-18}-10^{-17}$\,s\,s$^{-1}$ (\cite[Gotthelf \etal\ (2013)]{2013ApJ...765...58G}). The values of $P$ and $\dot{P}$ yield a characteristic age of a few $10^8$\,yr, many orders of magnitude larger than the age of the host supernova remnants ($0.3-7$\,kyr) indicating that CCOs were either born spinning nearly at their present periods or had an atypical magnetic field evolution. The timing inferred magnetic field at the surface is of the order of $10^{10}-10^{11}$\,G. Given this low $B$-field value, CCOs were interpreted as anti-magnetars, i.e. NSs born having weak magnetic fields that have not been effectively amplified through dynamo effects due to their slow rotation at birth. An alternative explanation, the so-called `hidden magnetic field' scenario, involves a fallback accretion episode of debris of the supernova explosion onto the newly born NS. Models showed that a typical magnetic field of a few $10^{12}$\,G can be buried in the crust by accreting a mass of $\sim 10^{-4}-10^{-2}$\,$M_{\odot}$ on timescale from hours to days. This process results in an external magnetic field lower than the internal hidden one, which might re-emerge on timescale of $10^3 - 10^5$\,yr once the accretion stops (see e.g., \cite[Torres-Forn{\'e} \etal\ (2016)]{2016MNRAS.456.3813T}).  

\begin{figure}[ht]
\centering
\includegraphics[width=0.9\columnwidth,trim={16cm 6cm 0cm 0cm},clip=true]{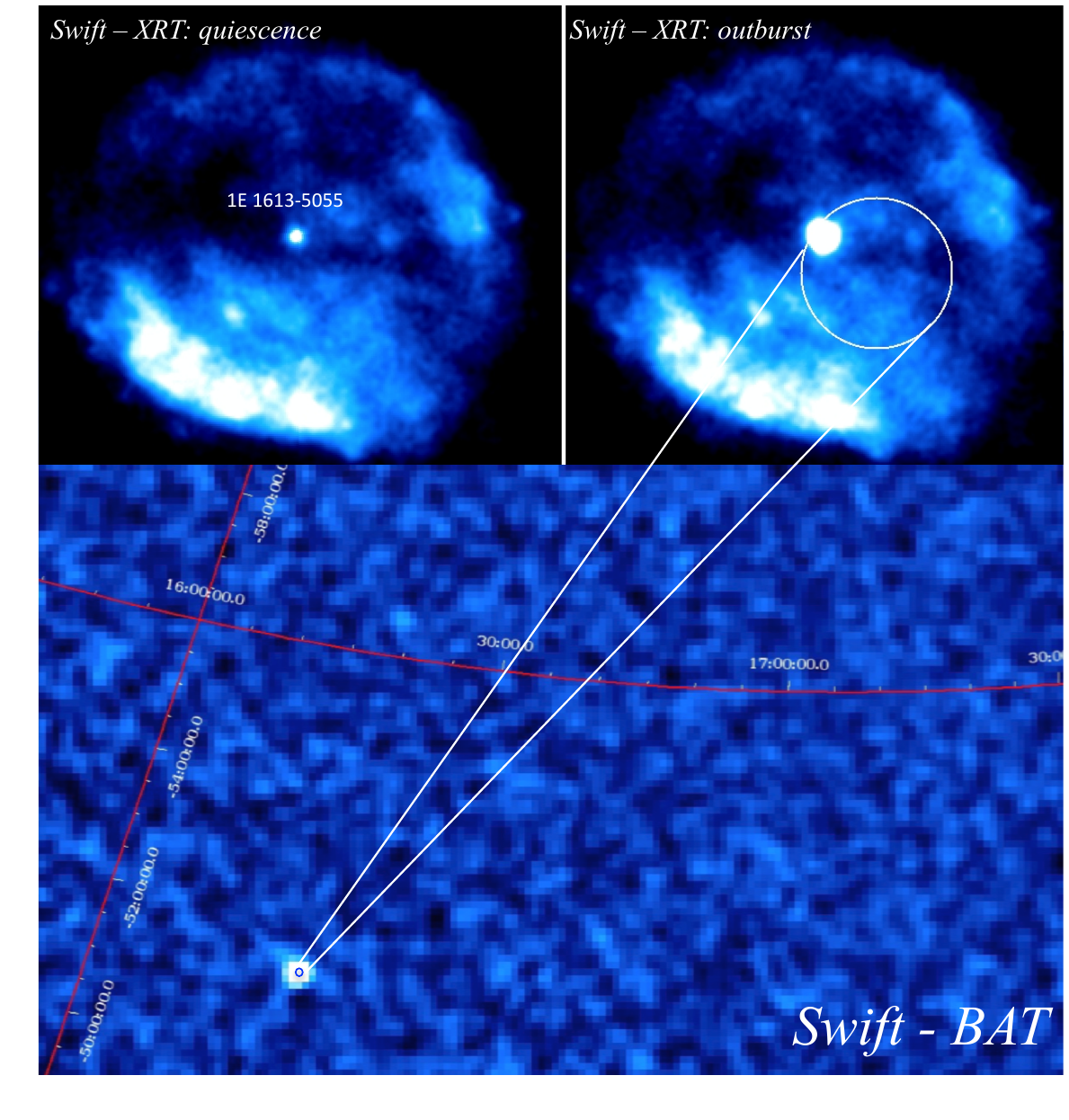}
\caption{{\it Swift}-BAT image of the burst detected on 2016 June 22 from the direction of the supernova remnant RCW\,103 (bottom). Two {\it Swift}-XRT co-added 1 -- 10\,keV images of RCW\,103 during the quiescent state of 1E\,161348$-$5055 (from 2011 April 18 to 2016 May 16) and in outburst (from 2016 June 22 to July 20). The white circle is the positional accuracy of the detected burst, which has a radius of 1.5 arcmin. Image from \cite[Rea \etal\ (2016)]{2016ApJ...828L..13R}.}
\label{fig:cco_outburst}
\end{figure}

{\it \underline{A unique CCO}}: 
Given its location at the centre of the 2-kyr-old supernova remnant RCW\,103, the soft thermal X-ray emission and the absence of a radio counterpart, 1E\,161348-5055 (1E\,1613) was classified as a CCO. It was considered a prototype of this class until observations revealed remarkable features that made 1E\,1613 stand out from the other CCOs. Unlike the other members of this group, it shows a strong flux variability on a monthly/yearly timescale, undergoing an outburst at the end of 1999 with a flux increase by a factor of $\sim 100$. Moreover, a long {\it XMM-Newton} observation carried out in 2005 has revealed a periodic modulation of 6.67\,hr (\cite[De Luca \etal\ (2006)]{2006Sci...313..814D}). Based on these characteristics, 1E\,1613 could be either the first low-mass X-ray binary in a supernova remnants with an orbital period of 6.67\,hr or a young isolated magnetar with a spin period of 6.67\,hr. In 2016, a new event shed light on the nature of this source. On 2016 June 22, a short ($\sim$10\,ms) magnetar-like burst was detected from the direction of the supernova remnant RCW\,103 and, meanwhile, an enhancement of the persistent flux by a factor of $\sim$100 with respect to the quiescent level was measured (see Figure\,\ref{fig:cco_outburst}). Follow-up observations confirmed that 1E\,1613 had entered a new magnetar-like outburst (\cite[D'A{\`\i} \etal\ (2016)]{2016MNRAS.463.2394D}, \cite[Rea \etal\ (2016)]{2016ApJ...828L..13R}). A hard X-ray,non-thermal spectral component was detected up to $\sim 30$\,keV for the first time and was modelled by a power law with photon index $\Gamma \sim 1.2$; while the soft X-ray spectrum was well fit using a double-blackbody model. The light curve displayed two broad peaks per cycle, shape dramatically different from the sinusoidal shape seen during the quiescent state. All the aspects caught during this event point towards a magnetar interpretation of 1E\,1613. However, its long periodicity makes this object unique among the known magnetars, whose rotational periods are smaller by three orders of magnitude. A very efficient braking mechanism is required to slow down the source in $\sim$2\,kyr from its period at birth to the currently measured value. Most models consider a propeller interaction with a fall-back disk that can provide an additional spin-down torque besides that due to dipole radiation (see e.g., \cite[Ho \& Andersson (2017)]{2017MNRAS.464L..65H}).

\section{Conclusions}

In this proceeding, I review the main groups of isolated NSs: magnetars, X-ray dim isolated neutron stars, rotation-powered pulsars and central compact objects. I showed that magnetar-like activity occurs in isolated neutron stars with a wide range of magnetic field much wider than we previously thought. These results blur the boundaries between different sub-classes and hint at an evolutionary link among them. We are a step closer to the grand unification of the neutron star zoo.

\end{document}